# Co-GISAXS as a New Technique to Investigate Surface Growth Dynamics

Meliha G. Rainville [‡], Christa Hoskin [‡], Jeffrey G. Ulbrandt [†], Suresh Narayanan [§], Alec R. Sandy [§], Hua Zhou [§], Randall L. Headrick [†] and Karl F. Ludwig, Jr. [∥][‡][*]

[‡] *Division of Materials Science and Engineering, Boston University, Boston, Massachusetts 02215 USA*

[∥] *Department of Physics, Boston University, Boston, Massachusetts 02215 USA*

[†] *Department of Physics and Materials Science Program, University of Vermont, Burlington, Vermont 05405 USA*

[§] *Advanced Photon Source, Argonne National Lab, Argonne, IL, 60439 USA*

[*]Email: ludwig@bu.edu

**Abstract**

Detailed quantitative measurement of surface dynamics during thin film growth is a major experimental challenge. Here X-ray Photon Correlation Spectroscopy with coherent hard X-rays is used in a Grazing-Incidence Small-Angle X-ray Scattering (i.e. Co-GISAXS) geometry as a new tool to investigate nanoscale surface dynamics during sputter deposition of a-Si and a-WSi$_2$ thin films.  For both films, kinetic roughening during surface growth reaches a dynamic steady state at late times in which the intensity autocorrelation function $g_2(q,t)$ becomes stationary.  The $g_2(q,t)$ functions exhibit compressed exponential behavior at all wavenumbers studied. The overall dynamics are complex, but the most surface sensitive sections of the structure factor and correlation time exhibit power law behaviors consistent with dynamical scaling.

# I. Introduction

As the result of continued improvement in coherent flux from high brilliance synchrotrons and free-electron lasers, X-ray Photon Correlation Spectroscopy (XPCS) offers new possibilities of measuring local dynamic processes in equilibrium and nonequilibrium systems [1-15]. XPCS shares physical principles with other Photon Correlation Spectroscopy (PCS) techniques [1]. When coherent light illuminates any material with disorder (static or dynamic), it gives rise to a speckle pattern that depends on the phase differences of the scattered wave from different parts of the sample. As the measured system undergoes changes, the speckle intensities fluctuate in time. XPCS is based on measuring speckle correlation, typically via the intensity autocorrelation function $g_2(t)$ [1-6]. In this study we show that XPCS offers a powerful new way to probe local surface dynamic processes during thin film deposition using coherent x-rays in a Grazing Incidence Small Angle X-ray Scattering geometry, i.e. via Co-GISAXS. GISAXS has its power by being surface sensitive, non-destructive, and applicable to a wide range of growth and experimental environments [16]. Consequently the Co-GISAXS approach gives unprecedented ability to measure dynamic evolution of the surface as a function of length scale. While Co-GISAXS has been used to examine the dynamics of capillary waves and in polymer films [4], to the best of our knowledge, this is the first time that it is used to study fundamental surface dynamics during thin film growth.

The interpretation of the speckle correlation from a nonequilibrium growth system can in general be very complicated. Therefore this study carefully examines the late time dynamic process of kinetic roughening during amorphous thin film growth after the surface roughness reaches a dynamic steady state. Kinetic roughening is a ubiquitous process but, despite much discussion, the extent to which actual systems obey simple models remains controversial. To optimize the scattering signal for these proof-of-concept experiments, we have deliberately chosen growth conditions which lead to relatively rough surfaces. Room temperature deposition of amorphous silicon (a-Si) and amorphous tungsten disilicide ($WSi_2$) through DC magnetron sputtering onto silicon (Si) and silicon dioxide ($SiO_2$) substrates respectively provides the basic growth environment in which crystallinity, grain boundaries and lattice mismatch with the substrate should have no impact. However, for growth at room temperature adatoms have limited surface mobility, resulting in complex surface and internal structures.

## II. Background

In the GISAXS experiment geometry (see Fig. 1) we take the z-direction to be along the sample normal, the x-direction to be the projected direction of the incident beam onto the sample plane and the y-direction to be the perpendicular to the x-direction in the sample plane. The measured wavevector transfer decomposes into two components: perpendicular to and parallel with the surface, $q_z$ and $q_{||}$ respectively. The $q_{||}$ includes both $q_x$ and $q_y$ components. However, since $q_x \ll q_y$ as a result of the small incidence and exit angles, and because the surfaces are isotropic, $q_y$ can be approximated as simply $q_{||}$.

The scattered x-rays are modeled by applying first order perturbation theory to the incident beam where the Born Approximation (BA) is valid. In other words, the intensity of the scattered beam is proportional to the square modulus of the Fourier Transform (FT) of the electron density in the material [17]. In the GISAXS regime, the Born Approximation for the x-ray scattering simplifies to the square modulus of the FT of the surface height (i.e. the height – height structure factor $S(q_x,q_y)$ in 2-D reciprocal space) in the low roughness limit where $q_z h(x,y) \ll 1$. When the incident or exit beam is near or below the critical angle for total reflection, the Distorted-Wave Born Approximation provides a more accurate description of the scattering. On a disordered surface, however, the scattering remains proportional to the height-height structure factor in the limit that $q_z' h(x,y) \ll 1$, where $q_z'$ is the z-component of the change in the scattered wavevector *inside* the material [18].

In general XPCS experiments can be run in homodyne or heterodyne modes. In homodyne experiments the intensity fluctuation of the scattered x-rays from the feature of interest alone is measured. On the other hand, in heterodyne experiments the scattered beam is made to interfere with a static or quasi-static reference and the intensity fluctuations of the resulting beam are studied [19]. Under conditions in which significant scattering from the bulk film is observed, we have discovered that heterodyning can occur between the bulk and surface signals during Co-GISAXS studies of thin film growth [20]. However, we focus here on experimental conditions giving homodyne behavior. The quantity typically evaluated in XPCS studies is the intensity autocorrelation function:

$$g_2(\boldsymbol{q},t) = \frac{\langle I(\boldsymbol{q},t')I(\boldsymbol{q},t'+t)\rangle}{\langle I(\boldsymbol{q},t')\rangle^2} \qquad (1)$$

where *I(q,t')* is the intensity at time *t'* at wavevector *q*. Angle brackets indicate a time averaging over *t'*. Scattered intensity is a second-order function of the electric fields and consequently $g_2(\mathbf{q},t)$ is fourth order in the fields. The electric fields are proportional to the FT of electron density. In a system with a scattered electric field that is a Gaussian random variable having zero mean, $g_2(\mathbf{q},t)$ can be decomposed into a simpler product of the autocorrelation function of the scattered electric field as given by the Siegert relation [1,4]:

$$g_2(\mathbf{q},t) = 1 + \beta(\mathbf{q})|g_1(\mathbf{q},t)|^2 \qquad (2)$$

where:

$$g_1(\mathbf{q},t) = \frac{\langle E(q,t')E(q,t'+t)\rangle}{\langle E(q,t')\rangle^2}. \qquad (3)$$

*β(q)* is a contrast term with a value between zero and one which depends on the experimental setup and the coherence of the incidence beam.

It is often reported that the $g_2(\mathbf{q},t)$ function can be well fit with a Kohlrausch-Williams-Watts form [21]:

$$g_2(\mathbf{q},t) = 1 + \beta(\mathbf{q})e^{-(t/\tau(\mathbf{q}))^n} \qquad (4)$$

where *τ(q)* is the *q*-dependent correlation time and *n* is an exponent that is specific to a materials process. If the system dynamics obeys a linear theory *n* takes a value of one, so that $g_2(q,t)$ becomes a simple exponential function. An example is a simple diffusive system where individual atoms undergo Brownian Motion. In this specific case of Fickian diffusion, the correlation time function is $\tau(\mathbf{q}) = 1/Dq^2$ where *D* is the diffusion constant[1-4]. If the exponent *n* takes a value larger or smaller than one, then dynamic processes cannot be explained by simple linear theory. The system then exhibits stretched exponential ($n < 1$) or compressed exponential ($n > 1$) behavior [22].

Room temperature deposition via DC magnetron sputter deposition leads to nonequilibrium growth dynamics where surfaces lack thermal energy to restructure themselves to find the lowest energy configuration. However, surfaces still go through local relaxation mechanisms that presumably depend on details of the local environment such as the curvature of the surface, leading to correlated surface growth. Following the initial stages of growth and increasing roughness surface correlations typically saturate at some cross-over time $t_\times$ so that roughening mechanisms become balanced by smoothening processes. Kinetic roughening is often discussed through dynamical scaling relationships which connect spatial and temporal correlations and are independent of many system details. A key surface growth scaling relation is the Family-Vicsek [23,24] scaling equation:

$$w(L,t) \sim L^\alpha f\left(\frac{t}{L^z}\right) \quad (5)$$

where $w(L,t)$ is the roughness of the interface or interface width, $L$ is the lateral length scale, $z$ is the dynamic growth exponent and $\alpha$ is the roughness exponent. $f\left(\frac{t}{L^z}\right)$ is a scaling function. For $u \ll 1$, $f(u)$ behaves as a power law $f(u) \to u^\beta$, and for $u \to \infty$, the scaling function approaches a constant value so that $w(L,t) \sim L^\alpha$. Therefore the surface width approaches a steady state value within the range of length scales studied. The crossover time between power law growth to a constant roughness scales with lateral length scale: $t_\times \sim L^z$.

Within the Family-Vicsek scaling relation, when evolution of the surface structure reaches a dynamical steady state the structure factor behaves as a power law: $S(q_\parallel) \sim q_\parallel^{-m}$. Since the structure factor is directly proportional to the square of the interface width, $m$ is related to $\alpha$ as [25]: $m = 2 + 2\alpha$. Additionally the autocorrelation function of surface heights can be related to the dynamic exponent $z$ [26]:

$$< h(q, t_1) h(q, t_2) > \sim g_{ss}(q^z |t_1 - t_2|) \quad (6)$$

By solving Eq (4) using Eq (6), the correlation time $\tau(q_\parallel)$ is found to be related to length scale $L$, or equivalently to wavenumber $q \sim 2\pi/L$, as $\tau(q_\parallel) \sim q_\parallel^{-z}$. Therefore the dynamic scaling exponent $z$ can be extracted directly from Co-GISAXS data under steady-state growth conditions. The ability to extract both $\alpha$ and $z$ from the same data set is very powerful. Since the remaining scaling exponent $\beta$ can also be recovered from $\beta = \alpha/z$, Co-GISAXS can be used to fully characterize the dynamics of a growing surface.

One of the best-known surface growth models is the Edward-Wilkinson [27,28] growth equation (EW). EW is often used to model random deposition with surface relaxation:

$$\frac{\partial h(x,t)}{\partial t} = \nu \nabla^2 h + \mu(x,t) \quad (7)$$

This temporal evolution of the surface height can be explained as the result of a surface tension "$\nu$" times the curvature of the surface height "$\nabla^2 h$" plus the random deposition noise $\mu(x,t)$. Deposition noise is usually modeled as Gaussian with average equal to zero. The scaling exponents for the EW model are $\alpha=0$ and $z=2$. The surface correlations exhibit exponential growth or exponential relaxation depending on the sign of the surface tension.

Kardar, Parisi and Zhang [25,29] (KPZ) suggested including the first nonlinear extension of the EW equation to have a more comprehensive growth equation that accounts for lateral growth. After adding the nonlinear correction term $\sqrt{1+(\nabla h)^2}$, which simplifies to $(\nabla h)^2$ in the limit of $|\nabla h| \ll 1$, to the EW model, the KPZ equation is

$$\frac{\partial h(x,t)}{\partial t} = \nu \nabla^2 h + \frac{\lambda}{2}(\nabla h)^2 + \eta(x,t) \quad (8)$$

Surface correlations reach saturation at a level determined partially by the nonlinear term $(\nabla h)^2$, which has $\lambda$ as a coefficient. There is no exact solution for scaling exponents for the KPZ equation in dimensions beyond 1+1 but many simulations and mathematical models have been used to prediction the exponents. For a 2+1 dimensional system, accepted values of $\alpha$ and $z$ from the literature are $\alpha \cong 0.4$ and $z \cong 1.6$. The $(\nabla h)^2$ nonlinear term determines the scaling exponents at long times and long wavelengths even if additional linear terms or nonlinear terms such as $\nabla^2(\nabla h)^2$ are added to Eq. 8. Indeed, a key attribute of the KPZ scaling is that more sophisticated growth models, such as ballistic growth models, exhibit similar scaling at long length scales and times.

## III. Experimental

Real-time x-ray scattering studies are performed in a custom-built ultrahigh vacuum (UHV) chamber with a base pressure of $3\times10^{-8}$ Torr capable of holding a DC magnetron sputter deposition source. The deposition chamber is installed onto a diffractometer on beamline 8-ID-I of the Advanced Photon Source (APS) located at Argonne National Laboratory. Incoming partially coherent x-rays with 7.38 keV photon energy are focused to a beam of dimension of $20 \times 4$ μm$^2$ at the sample position to improve speckle contrast. Grazing incidence angles of x-ray beam are chosen to be less than or equal to the critical angle of total external reflection ($\alpha_c$) of the deposited materials to decrease the bulk scattering and to improve surface sensitivity. A two-dimensional Princeton Instruments direct illumination CCD camera, which is located 4067 mm away from the sample, is set to measure the scattered intensity with two-second intervals with a readout time of 1 second and pixel size of 20 × 20 µm². In order to record a wider region of $q_{||}$ space, the detector location is periodically moved horizontally while keeping the detector-to-sample distance constant. Each detector location shares 20 mm of overlap with the previous one to guarantee continuity of the data. The scattered x-rays are recorded around the Yoneda wing position [30], which is where enhanced surface scattering occurs when the exit angle of the scattered x-rays $\alpha_{out} = \alpha_c$. When $\alpha_{out}$ is higher than $\alpha_c$, the scattering becomes less surface sensitive and starts to have more bulk scattering component. On the other hand, the scattering becomes more surface sensitive when $\alpha_{out} < \alpha_c$ but less intense. In order to the check the effect of exit angle on $S(q_{||})$ and $g_2(q_{||}, t)$, the recorded data is analyzed at three different $q_z$ locations: 0.1° above the exit critical angle, 0.1° below the exit critical angle and at the Yoneda wing position itself (i.e. at the exit critical angle).

The temporal evolution of scattered intensity was used to determine when the surface roughening process reached a steady state. In general, the scattering at higher wavenumbers saturates sooner than at smaller ones. The steady state conditions for all length scales examined were reached within 8000 seconds after the deposition started. All other data presented in this study only includes results taken after steady state conditions were satisfied.

The deposition of a-Si and a-WSi$_2$ thin films is performed using DC magnetron sputtering at room temperature. Argon gas of 99.999% purity is used for the plasma. The sputtering targets are pre-sputtered for an hour with shutter closed to remove any contamination and oxide layers before deposition starts. The substrates have 1 × 2 cm$^2$ dimensions and are solution cleaned before being put into the vacuum chamber. The a-Si thin films are deposited on the 600$\mu m$ thick Si (111) wafers with Ar gas pressure of 10 mTorr. Two different deposition powers (20W and 40W) are used to investigate effects of the deposition rate on surface dynamics. The a-WSi$_2$ thin films are grown onto 200$\mu m$ thick SiO$_2$ templates with 25W deposition power and with 10mTorr Ar gas pressure.

Post-growth specular X-ray reflectivity investigations of the a-Si and a-WSi$_2$ thin films are performed to measure $\alpha_c$ and the density of the films after each deposition is completed. The critical angle of the a-Si thin films is measured as 0.21° which is 0.03° less than the critical angle of crystalline Si at this energy. The calculated density of the a-Si thin films using these critical angle measurements as well as *ex-situ* SEM micrographs and microbalance results, suggests that the grown films have 70% of the density of crystalline silicon. The measured critical angle for a-WSi$_2$ thin films is the same as expected for crystalline WSi$_2$, 0.45°, suggesting that the films have the same density as crystalline WSi$_2$.

## IV. Results
### IV.A: a-Si Thin Film Deposition
During the Co-GISAXS measurements, the incidence angle for incoming x-rays is set to

0.16° which is well below the $\alpha_c$ of the films to emphasize scattering from the surface and near-surface (< 5 nm) layers. The values of in-plane reciprocal space accessed were 0.005 Å$^{-1}$ < $q_{||}$ < 0.121 Å$^{-1}$, corresponding to lateral length scales of $2\pi/q_{||}$ ~ 50-1250 Å. Exit angles measured on the area detector were 0.028° < $\alpha_{out}$ < 0.394°.

Figure 2 shows the GISAXS intensity, which is proportional to the structure factor $S(q_{||})$, measured after the surface roughness evolution reached a steady state. The structure factors measured at the three distinct exit angles all behave as a power law at low $q_{||}$ but there is increased scattering with a shoulder at the higher wavenumbers. The amount of increased scattering at high $q_{||}$ and the exact exponent of the power law at low $q_{||}$ change with exit angle. The more pronounced bump at higher exit angle suggests that the extra scattering is coming from the near-surface layers.

All structure factors are fit by a heuristic equation which is the sum of a Gaussian function and a power law

$$I(q_{||}) = I_p q_{||}^{-m} + I_g e^{-q^2/2\sigma^2} \qquad (9)$$

The fit results for each structure factor (above, at, and below the Yoneda wing) can be found in Table 1. The results of the fits are generally consistent, though the exponent of the power law increases slightly as the exit angle increases. The width of the Gaussian function indicates that near-surface scattering is coming from structures approximately 100 Å in lateral size. Figure 3 shows the ratio of the power-law to Gaussian components at $q_{||}$ = 0.02 Å$^{-1}$ as a function of exit angle. It's seen that the power-law component increases rapidly relative to the Gaussian component as the exit angle decreases below the critical angle. This suggests that the power-law component of the scattering comes from the surface itself while the Gaussian component comes from the near-surface region.

| Label | Incidence Angle | Exit Angle | Power Law Exponent (± 0.25) | Gaussian Width (σ) | Correlation size (2π/σ) |
|---|---|---|---|---|---|
| Below Yoneda | 0.16° | 0.11° | 2.45 | 0.061 Å$^{-1}$ | 103 Å |
| Yoneda | 0.16° | 0.21° | 2.72 | 0.054 Å$^{-1}$ | 116 Å |
| Above Yoneda | 0.16° | 0.31° | 2.90 | 0.068 Å$^{-1}$ | 92 Å |

Table 1. Parameters from fits of Eq. 9 to the structure factors of a-Si thin films during steady state growth.

After the scattering reaches a steady state, the dynamics are investigated through the intensity autocorrelation function $g_2(q_\parallel,t)$. Since scattering at exit angles above the Yoneda peak shows increased contributions from near-surface scattering which can lead to interference between surface and near-surface scattered waves [20], we focus on the scattered intensity at the Yoneda wing and below it. The $g_2(q_\parallel,t)$ results are fitted with Eq. 4 to yield the correlation times $\tau(q_\parallel)$ and exponents $n(q_\parallel)$. As Fig. 4 shows, the $g_2(q_\parallel,t)$ functions clearly show compressed exponential behavior. The fit correlation times are presented in Fig. 5. At long time scales the beamline optics may not be stable, so the longest correlation times should be interpreted cautiously. At both the Yoneda wing location and below, $\tau(q_\parallel)$ decreases approximately as a power law and then decreases more slowly at larger $q_\parallel$ (i.e. for real-space correlations < 80 *nm*). The modest $\tau(q_\parallel)$ regions displaying power law behavior are fit and the resulting exponents are $z \sim$ 1.24 at the Yoneda position and $z \sim 1.05$ for the exit angle below it.

Figure 6 shows the measured exponents $n(q_\parallel)$ from the $g_2(q_\parallel,t)$ fits as a function of wavenumber. Their behavior is complex. In general the compressed exponents stay less than 1.5 for $q_\parallel < 0.02 Å^{-1}$ at all exit angles but then increase to approximately 2.

In order to investigate the effects of the deposition rate on the surface dynamics, the deposition power is increased from 20W to 40W and the GISAXS scattering is examined at a single detector position. Deposition studies have shown that the deposition rate is approximately linearly proportional to deposition power. As the deposition rate is doubled, the structure factor remains unchanged (Fig. 7a). On the other hand, the values of $\tau(q_\parallel)$ decrease by the factor of 1.8 at a given wavenumber as the deposition rate is doubled (Fig. 7b). This confirms that the time scales for dynamics at the surface are only driven by the deposition itself, not by equilibrium thermal effects. We have also found that the surface dynamics cease entirely when the deposition is halted (not shown). The compressed exponents remain unchanged (Fig. 7c).

*Ex-situ* cross-sectional SEM study of the a-Si thin film shows highly elongated structural domains [31] within the film that are aligned parallel to the surface normal (Fig. 8). Each domain has a width of approximately 3000 Å and a height that can be as large as the total film thickness. The domains are separated from each other by narrow, deep valleys. In contrast, the Gaussian-components of the x-ray results are the result of near-surface structures with only a $100Å$ size scale. Though it is more difficult to see these finer structures from the SEM image, the existence of finer structures within these structural domains has been reported in the literature [32,33]. Therefore, it seems likely that the 3000Å wide structural domains observed in SEM are formed of smaller structures, which cause the near-surface x-ray scattering observed.

**IV.B: a-WSi$_2$ Thin Film Deposition**

The experimental geometry was chosen to enhance the surface sensitivity while maintaining sufficient signal-to noise-ratio. The incidence angle for incoming x-rays was set to 0.40°, which is lower than the critical angle for total external reflection for a-WSi$_2$ thin films, and the scattered x-rays were recorded at exit angles between 0.36° and 0.7°. The in-plane scattering was examined over a similar range as for the a-Si growth.

Figure 9 shows the GISAXS intensities of a-WSi$_2$ thin films after the surface growth reached a dynamic steady state. Similarly to the a-Si thin film results, all the intensities exhibit two regions: a power law region at low $q_{||}$ and a region of increased scattering with a shoulder at high $q_{||}$. The shape of the structure factor curves barely changes between different exit angles. As before, all the structure factors are fit by power-law and Gaussian components as given by Eq. 9; the fit results for each structure factor can be found in Table 2. As shown in Fig. 3, the ratio of power-law to Gaussian behavior increases sharply as the exit angle goes below the critical angle, again suggesting that the power-law behavior is associated with the surface itself and the Gaussian with the near-surface region. The exponent of the power law decreases slightly as the exit angle of the x-rays increases. The width of the Gaussian function suggests that near-surface scattering is coming from structures which are approximately 90 Å in lateral size.

| Label | Incidence Angle | Exit Angle | Power Law Exponent ($\pm 0.2$) | Gaussian Width ($\sigma$) | Correlation size ($2\pi/\sigma$) |
|---|---|---|---|---|---|
| Below Yoneda | 0.40° | 0.35° | 2.50 | 0.065 Å$^{-1}$ | 97 Å |
| Yoneda | 0.40° | 0.45° | 2.52 | 0.067 Å$^{-1}$ | 94 Å |
| Above Yoneda | 0.40° | 0.55° | 2.14 | 0.094 Å$^{-1}$ | 67 Å |

Table 2. Parameters from fits of Eq. 9 to the structure factors of a-WSi$_2$ thin films during steady state growth.

The local surface dynamics of the a-WSi$_2$ films was studied via the intensity autocorrelation function $g_2(q_\parallel,t)$, and correlation times $\tau(q_\parallel)$ and exponents $n(q_\parallel)$ were extracted similarly to the a-Si thin film case. Figure 10 shows how the correlation times depend on wavenumber for exit angles at the Yoneda wing and below. The stability of the beamline at long time scales (affecting the low $q_\parallel$ correlation times) and contribution from the near-surface scattering at high $q_\parallel$ caused $\tau(q_\parallel)$ to behave as a power law in very limited region for both exit angles. In this region $\tau(q_\parallel)$ varies as $\tau \sim q_\parallel^{-2.00}$ at the Yoneda wing and $\tau \sim q_\parallel^{-1.67}$ below it. The exponents $n(q_\parallel)$ obtained from fits of the $g_2(q_\parallel,t)$ function for a-WSi$_2$ films are plotted in Fig. 11; the compressed exponents are between 1.2 and 2, roughly comparable to what was found for the a-Si growth.

The deposited a-WSi$_2$ thin films were studied by *ex-situ* cross-sectional SEM (Fig. 12) to have a better understanding of structures within the film. Similar to the a-Si thin films, there are highly elongated structures within the a-WSi$_2$ thin films. By comparison to cross-sectional SEM images of a-Si thin films, it can be concluded that the structural domains in a-WSi$_2$ are narrower and still very tall. The finer structures (~200Å) are more pronounced than in the a-Si SEM image. The near-surface layer x-ray scattering is presumably from these finer structures sitting under the surface.

V. Discussion and Conclusions

The results for sputter deposited growth of a-Si and a-WSi$_2$ show similar systematic behaviors, allowing more general conclusions to be drawn. The x-ray scattering and SEM micrographs show that both film structures are complex. *Post-facto* AFM analysis shows that surface roughness is ~ 5 nm for the a-Si films and ~ 2 nm for the a-WSi$_2$ films. These are comparable to the sampling depth of the x-rays, so the results here should be considered as sampling the width of the film-vacuum interface. While surface scattering is consistent with power law spatial correlations on the longest length scales examined here, the structure at shorter length scales appears to be dominated by near-surface structures – possibly nano-columns that have been reported in earlier literature [32,33].

The Co-GISAXS technique has allowed us to examine the steady-state *dynamics* of kinetic roughening for the first time. Increasing the deposition rate shows that the dynamics are driven by the deposition process itself under the conditions studied here. However, just as the real-space structure of these films is complex, so is their dynamics. All $g_2(q_{||},t)$ functions exhibit compressed exponential relaxation, which is inconsistent with linear models such as EW. Compressed exponents have been previously measured in wide variety of soft materials [22,34-37] (gels, sponges, clays and emulsions), in magnetic and in electronic [38,39] systems. Simulations of the KPZ model show that nonlinearities can produce compressed exponents, and the exponents measured here could be indicative of the nonlinear surface growth dynamics [40]. However the particular wavenumber dependence of $n(q_∥)$ seen in these experiments is, to our knowledge, unique.

The measured correlation times are consistent with a power law behavior at the lower wavenumbers accessible but show a marked flattening toward the higher wavenumbers. This could be associated with the presence of near-surface structure seen in the scattered intensity itself. While simplified models such as KPZ may capture some of the essential physics of the surface growth dynamics over a limited range of length scales, additional mechanisms may be equally important at other length scales. In particular, continuum models such as KPZ make the basic assumption that the local surface growth velocity is uniquely determined through a specific function of the local surface gradient $\nabla h$. Such models neglect important interactions between surface and near surfaces features (e.g. through relaxation of strain), as well as other nonlocal effects such as shadowing.

Within structure factor and correlation time power law regimes, exponents measured here vary but are clearly inconsistent with predictions of the linear EW model. They are rather closer to those predicted by the nonlinear KPZ model, but a detailed understanding of the compressed exponents of the $g_2(q_∥,t)$ function predicted by the model does not yet exist. Now that such detailed experimental information about surface dynamics is available from Co-GISAXS, it's clear that a more detailed dialogue of experiment with theory/modelling modeling of amorphous growth is warranted.

## VI. Acknowledgments

We thank Ray Ziegler for beamline support and Alex DeMasi for experimental support. MR, KL, and CH were supported by the U.S. Department of Energy (DOE) Office of Science, Office of Basic Energy Sciences (BES) under DE-FG02-03ER46037; RH and JU were supported by DOE BES grant DE-FG02-07ER46380. This research used resources of the Advanced Photon Source (APS), a U.S. DOE Office of Science User Facility operated for the DOE Office of Science by Argonne National Laboratory under Contract No. DE-AC02-06CH11357.

## References

[1] B.J. Berne and R. Pecora, *Dynamic light scattering : with applications to chemistry, biology, and physics*. Dover edn, (Dover Publications, 2000).

[2] M. Sutton, C. R. Physique **9**, 657 (2008).

[3] F. Livet, Acta Crystallogr. Sect. A **63**, 87 (2007).

[4] S.K. Sinha, Z. Jiang and L.B. Lurio, Adv. Mater. **26**, 7764 (2014).

[5] M. Sutton, S.G.J. Mochrie, T. Greytak, S.E. Nagler, L.E. Berman, G.A. Held and G.B. Stephenson, Nature **352**, 608 (1991).

[6] O.G. Shpyrko, J. Synchrotron Rad. **21**, 1057 (2014).

[7] F. Zhang, A.J. Allen, L.E. Levine, J. Ilavsky and G.G. Long, Metall. Mater. Trans. A **43A**, 1445 (2012).

[8] C. Gutt, T. Ghaderi, V. Chamard, A. Madsen, T. Seydel, M. Tolan, M. Sprung, G. Grubel and S. K. Sinha, Phys. Rev. Lett. **91**,076104 (2003).

[9] T. Thurn-Albrecht, W. Steffen, A. Patkowski, G. Meier, E.W. Fischer, G. Grubel and D.L. Abernathy, Phys. Rev. Lett. **77**, 5437 (1996).

[10] G.B. Stephenson, A. Robert and G. Grubel, Nature Mater. **8**, 702 (2009).

[11] S. O. Hruszkewycz, M. Sutton, P.H. Fuoss, B. Adams, S. Rosenkranz, K.F. Ludwig, Jr., W. Roseker, D. Fritz, M. Cammarata, D. Zhu, S. Lee, H. Lemke, C. Gutt, A. Robert, G. Grubel and G.B. Stephenson, Phys. Rev. Lett. **109**, 185502 (2012).

[12] I. A. Vartanyants, D. Grigoriev and A.V. Zozulya, Thin Solid Films **515**, 5546 (2007).

[13] O. Bikondoa, D. Carbone, V. Chamard and T. H. Metzger, J. Phys.: Condens Matter **24**, 445006 (2012).

[14] H. Kim, A. Ruhm, L.B. Lurio, J.K. Basu, J. Lal, D. Lumma, S.G.J. Mochrie and S. K. Sinha, Phys. Rev. Lett. **90**, 068302 (2003)

[15] T. Seydel, A. Madsen, M. Tolan, G. Grubel and W. Press, Phys. Rev. B **63**, 073409 (2001).

[16] G. Renaud, R. Lazzari and F. Leroy, Surf. Sci. Rep. **64**, 255 (2009).

[17] J. Als-Nielsen, and D. McMorrow, *Elements of Modern X-ray Physics* (John Wiley and Sons, New York, 2001).

[18] S. K. Sinha, E.B. Sirota, S. Garoff and H.B. Stanley, Phys. Rev. B. **38**, 2297 (1988).

[19] W. H. de Jeu, A. Madsen, I. Sikharulidze and S. Sprunt, Physica B **357**, 39 (2005).

[20] J.G. Ulbrandt, M.G. Rainville, C. Hoskin, S. Narayanan, A.R. Sandy, H. Zhou, K.F. Ludwig, Jr. and R.L. Headrick, arXiv:1507.03694 , submitted to *Nature Physics*.

[21] W. Williams and D.C. Watts, Trans. Faraday Soc. **66**, 80 (1970).

[22] L. Cipelletti, L. Ramos, S. Manley, E. Pitard, D.A. Weitz, E.E. Pashkovski and M.

Johansson, Faraday Discuss. **123**, 237 (2003).

[23] F. Family and T. Vicsek, J. Phys. A-Math. Gen. **18**, 75 (1985).

[24] T. Vicsek and F. Family, Phys. Rev. Lett. **52**, 1669 (1984).

[25] A.L. Barabasi and H.E. Stanley, *Fractal Concepts in Surface Growth*. (Cambridge University Press, Cambridge, 1995), p. 305; p. 56.

[26] K. Sneppen, J. Krug, M.H. Jensen, C. Jayaprakash and T. Bohr, Phys. Rev. A **46**, 7351 (1992).

[27] F. Family, Physica A **168**, 561 (1990).

[28] S.F. Edwards and D.R. Wilkinson, P. Roy. Soc. A-Math. Phy. **381**, 17 (1982).

[29] M. Kardar, G. Parisi and Y.C. Zhang, Phys. Rev. Lett. **56**, 889 (1986).

[30] Y. Yoneda, Phys. Rev. **131**, 2010 (1963).

[31] J.A. Thornton, J. Vac. Sci. Technol. **11**, 666 (1974).

[32] R. Messier, A. P. Giri and R.A. Roy, J Vac. Sci. Technol. A **2**, 500 (1984).

[33] B. Marsen and K. Sattler, Phys. Rev. B **60**, 11593 (1999).

[34] R. Bandyopadhyay, D. Liang, H. Yardimci, D. A. Sessoms, M. A. Borthwick, S.G.J. Mochrie, J.L. Harden and R.L. Leheny, Phys. Rev. Lett **93**, 228302 (2004).

[35] H. Guo, J.N. Wilking, D. Liang, T. G. Mason, J. L. Harden and R. L. Leheny, Phys. Rev. E **75**, 041401 (2007).

[36] P. Falus, M.A. Borthwick, S. Narayanan, A.R. Sandy and S.G.J. Mochrie, Phys. Rev. Lett. **97**, 066102 (2006).

[37] L. Cipelletti and L. Ramos, J. Phys.: Condens. Matter. **17**, 253 (2005).

[38] O.G. Shpyrko, E.D. Isaac, J.M. Logan, Y.J. Feng, G. Aeppli, R. Jaramillo, H.C. Kim, T.F. Rosenbaum, P. Zschack, M. Sprung, S. Narayanan, and A.R. Sandy, Nature **447**, 68 (2007).

[39] S.W. Chen, H. Guo, K.A. Seu, K. Dumesnil, S. Roy and S.K. Sinha, Phys. Rev. Lett. **110**, 217201 (2013).

[40] M. Mokhtarzadeh, unpublished.

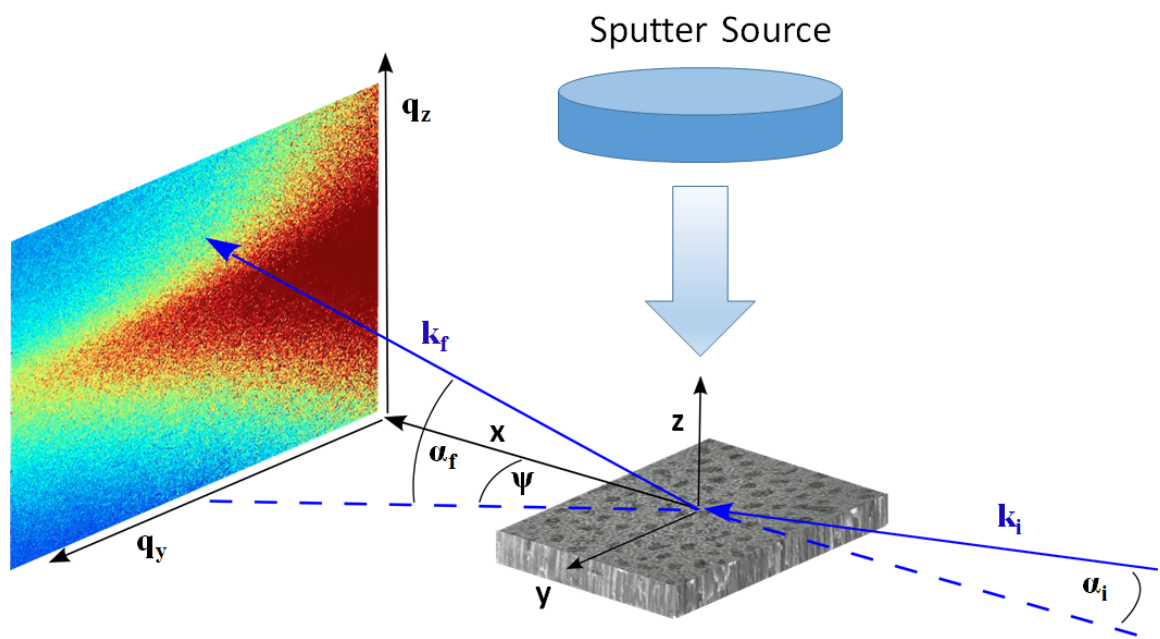

Figure 1. Schematic diagram of Co-GISAXS measurements during sputter deposition of Si and $WSi_2$.

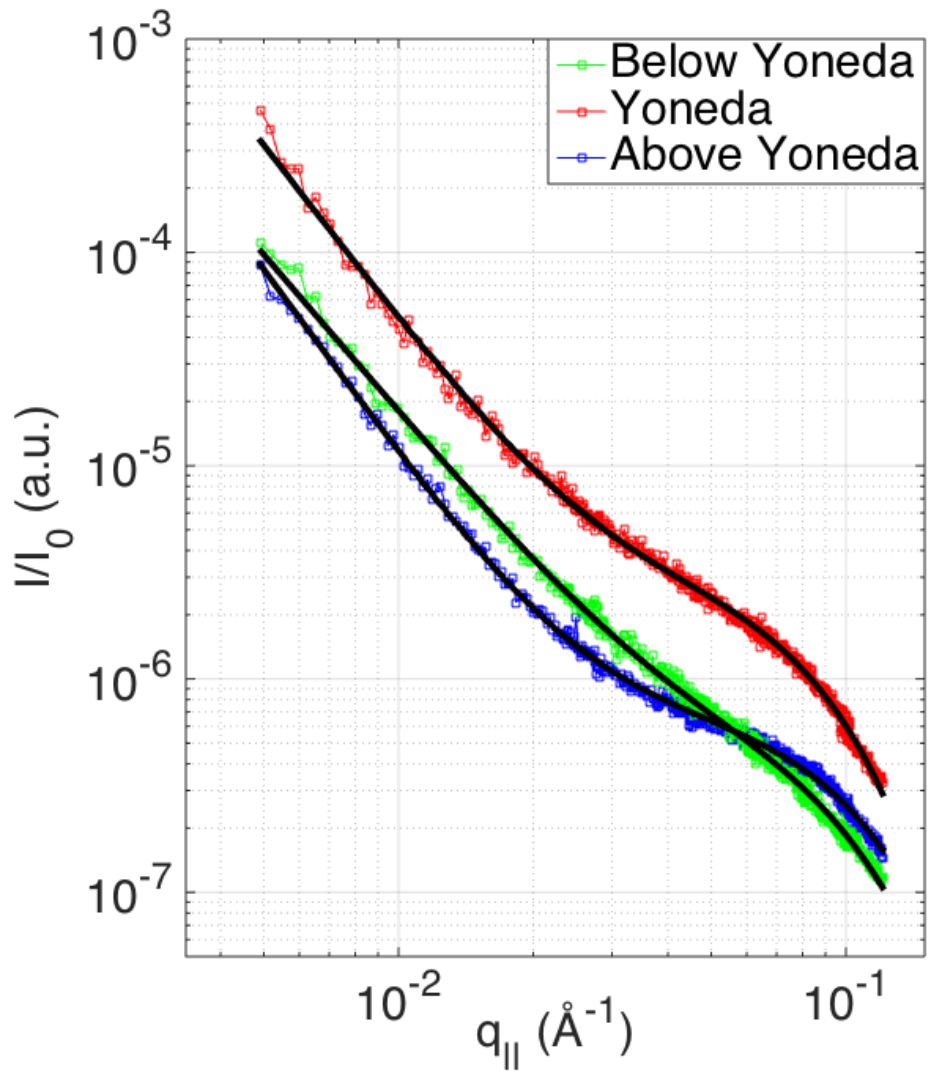

Figure 2. Steady-state GISAXS intensities measured at exit angles below, at, and above the Yoneda wing during a-Si thin film growth. The solid lines are fits to Eq. 9; fit parameters are given in Table 1.

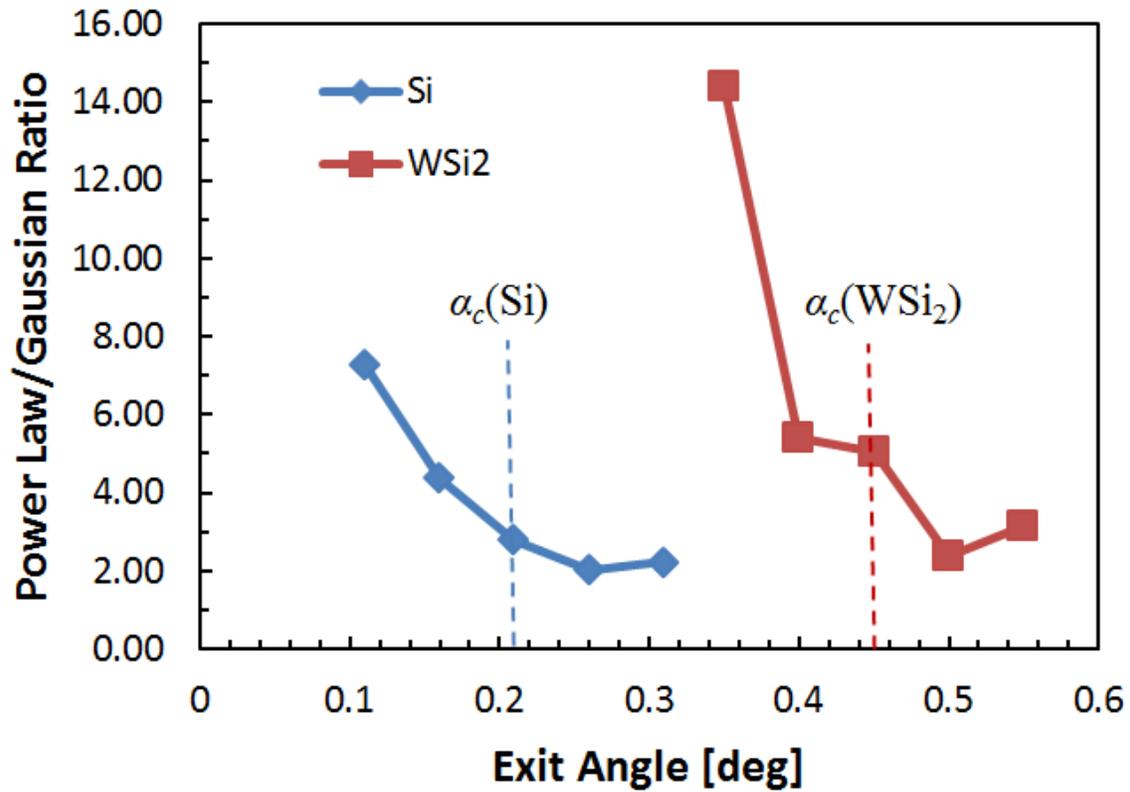

Figure 3. Ratio of Power Law to Gaussian component of the structure factor at $q_{//}$ = 0.02 Å$^{-1}$ as a function of exit angle for growth of a-Si and a-WSi$_2$. The ratio grows significantly below the critical angles for the two films.

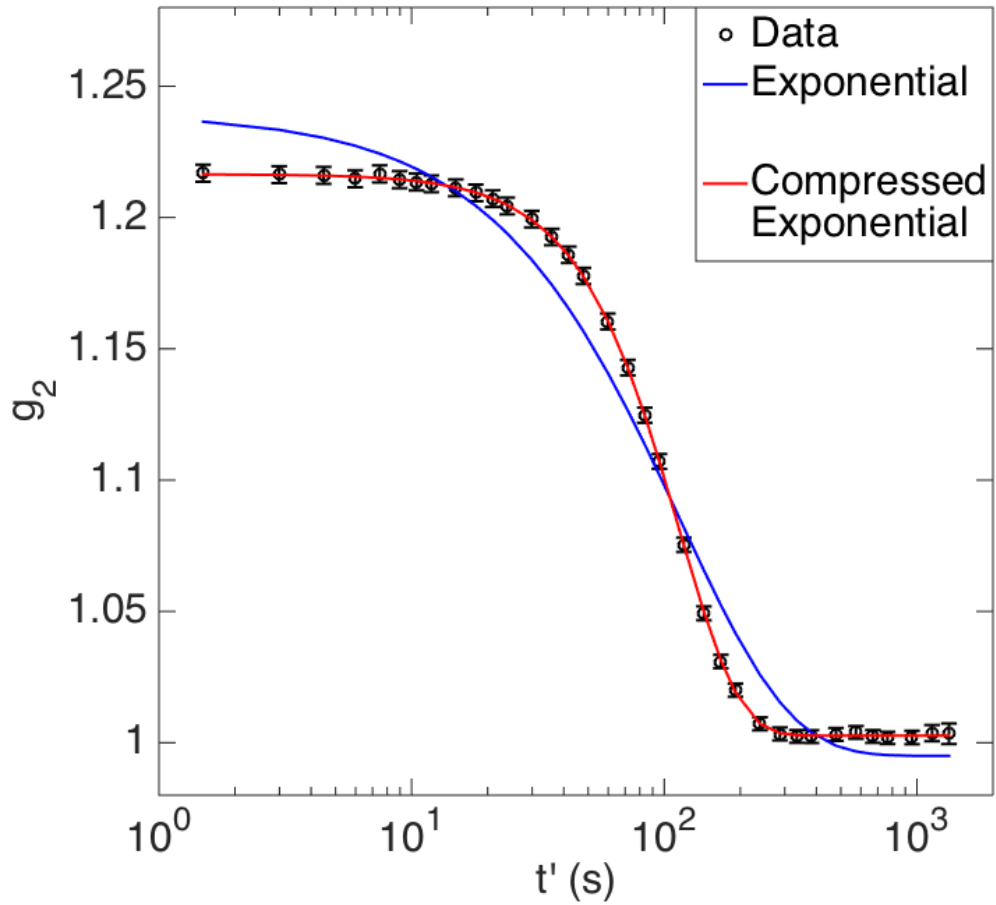

Figure 4. Typical homodyne $g_2(t)$ intensity auto-correlation function for steady-state growth of a-Si and a-WSi$_2$. The correlation decay follows a compressed exponential.

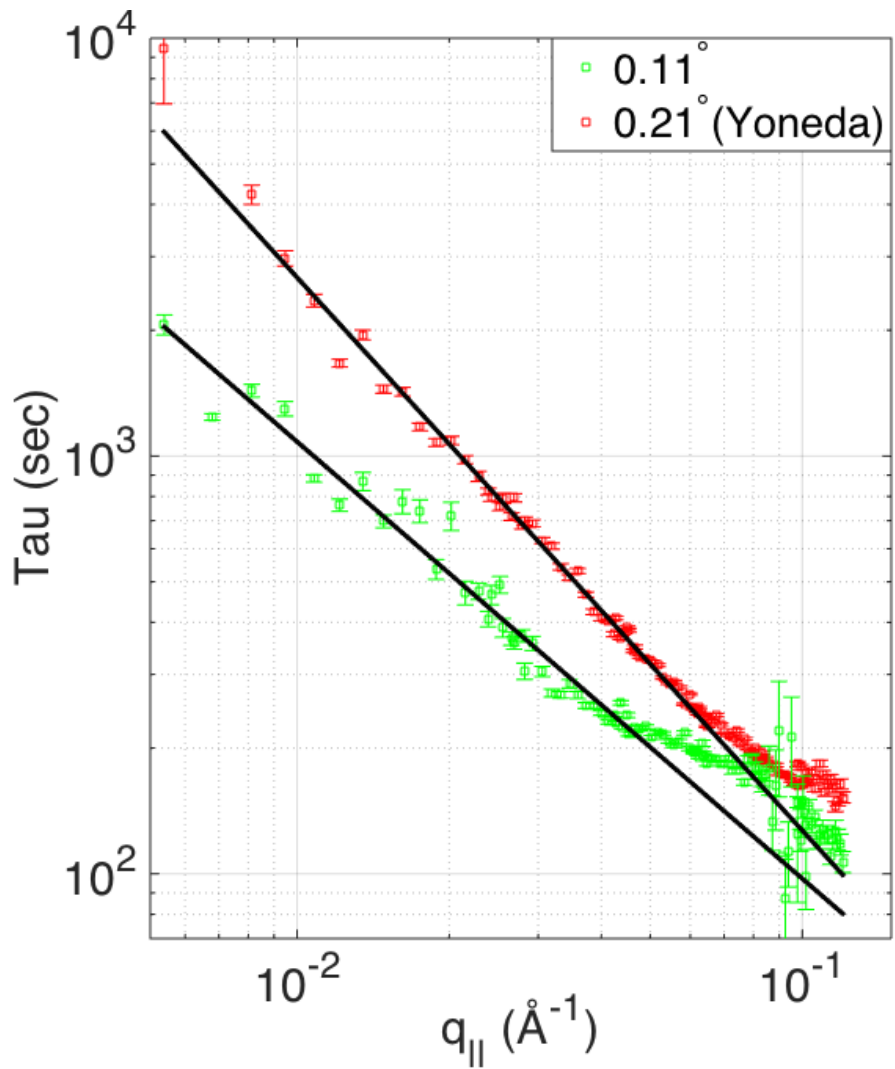

Figure 5. Correlation times for a-Si thin film growth measured at exit angles below and at the Yoneda wing.

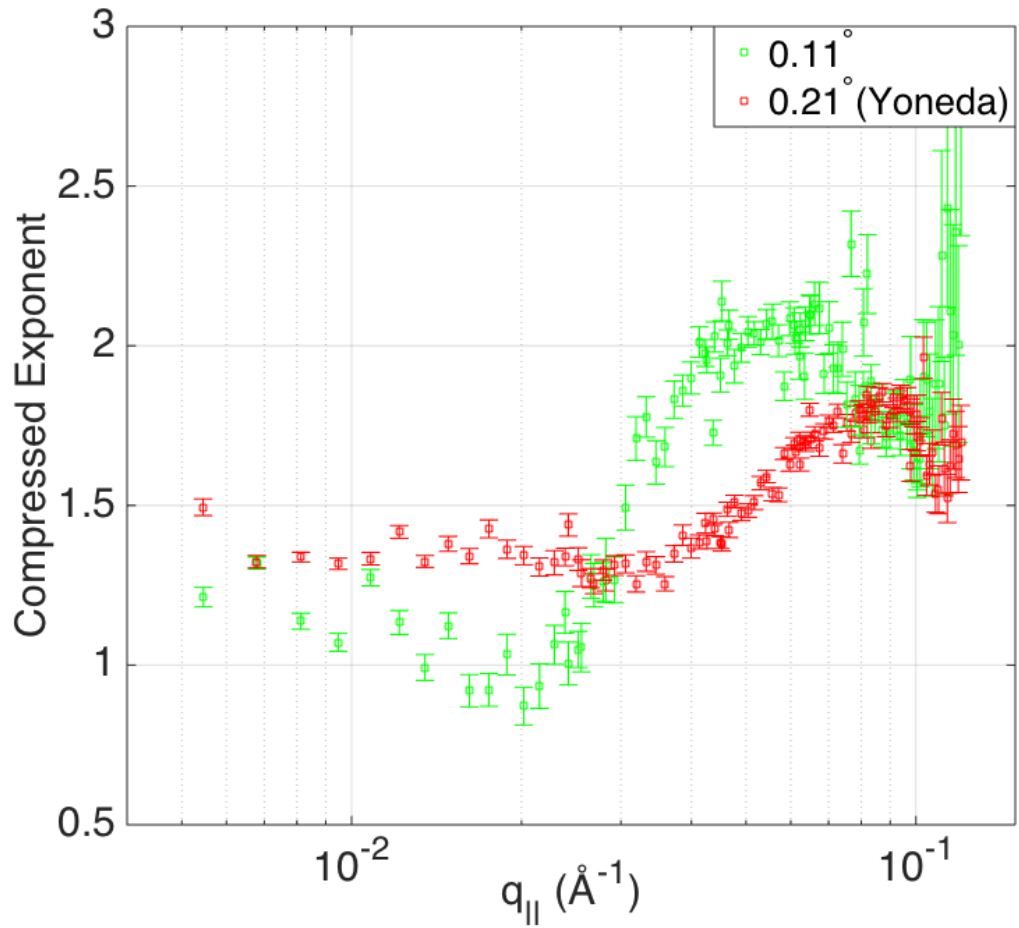

Figure 6. Compressed exponents from the $g_2(t)$ fits for exit angles below and at the Yoneda wing for a-Si growth.

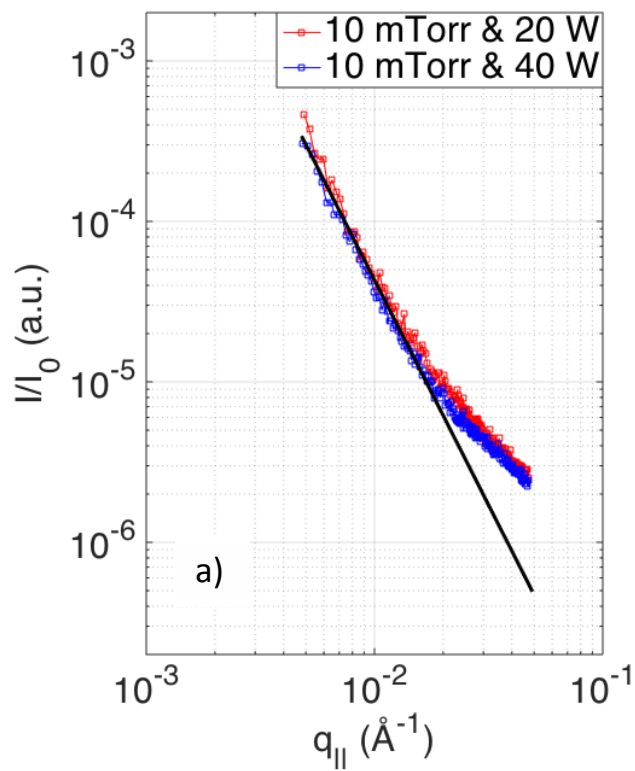
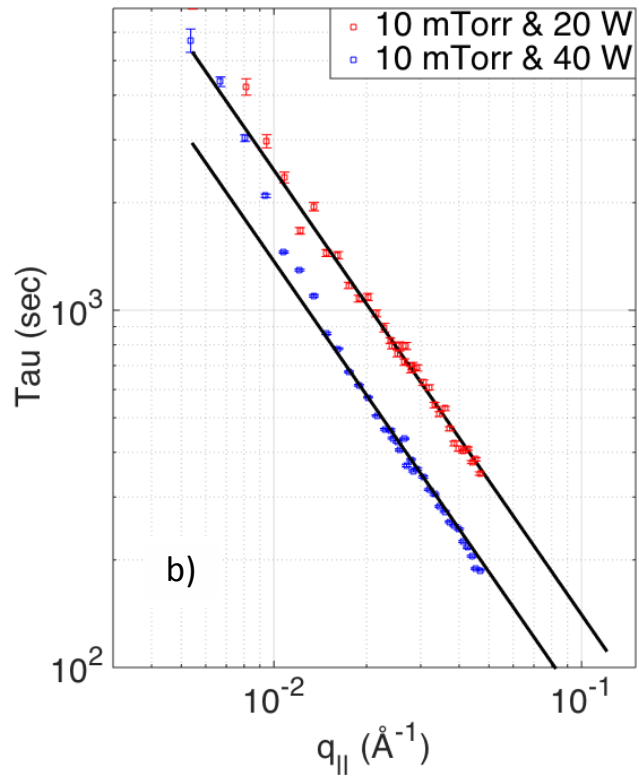
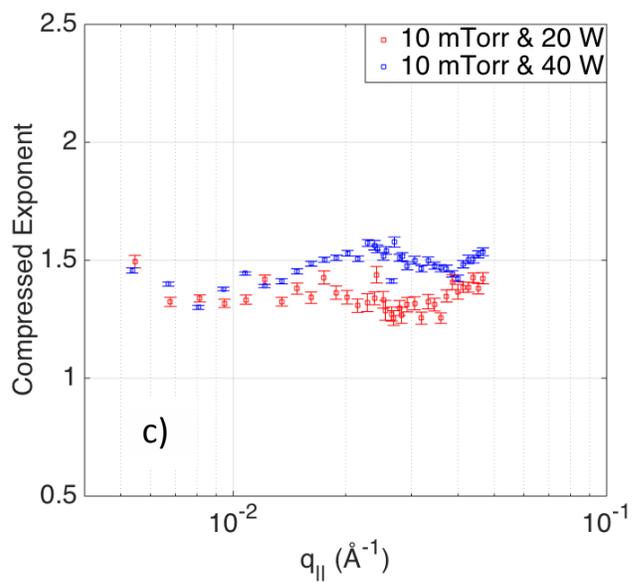

Figure 7. Comparison of a-Si deposition at 20W and 40W.
a) GISAXS intensity
b) correlation time and
c) compressed exponent.

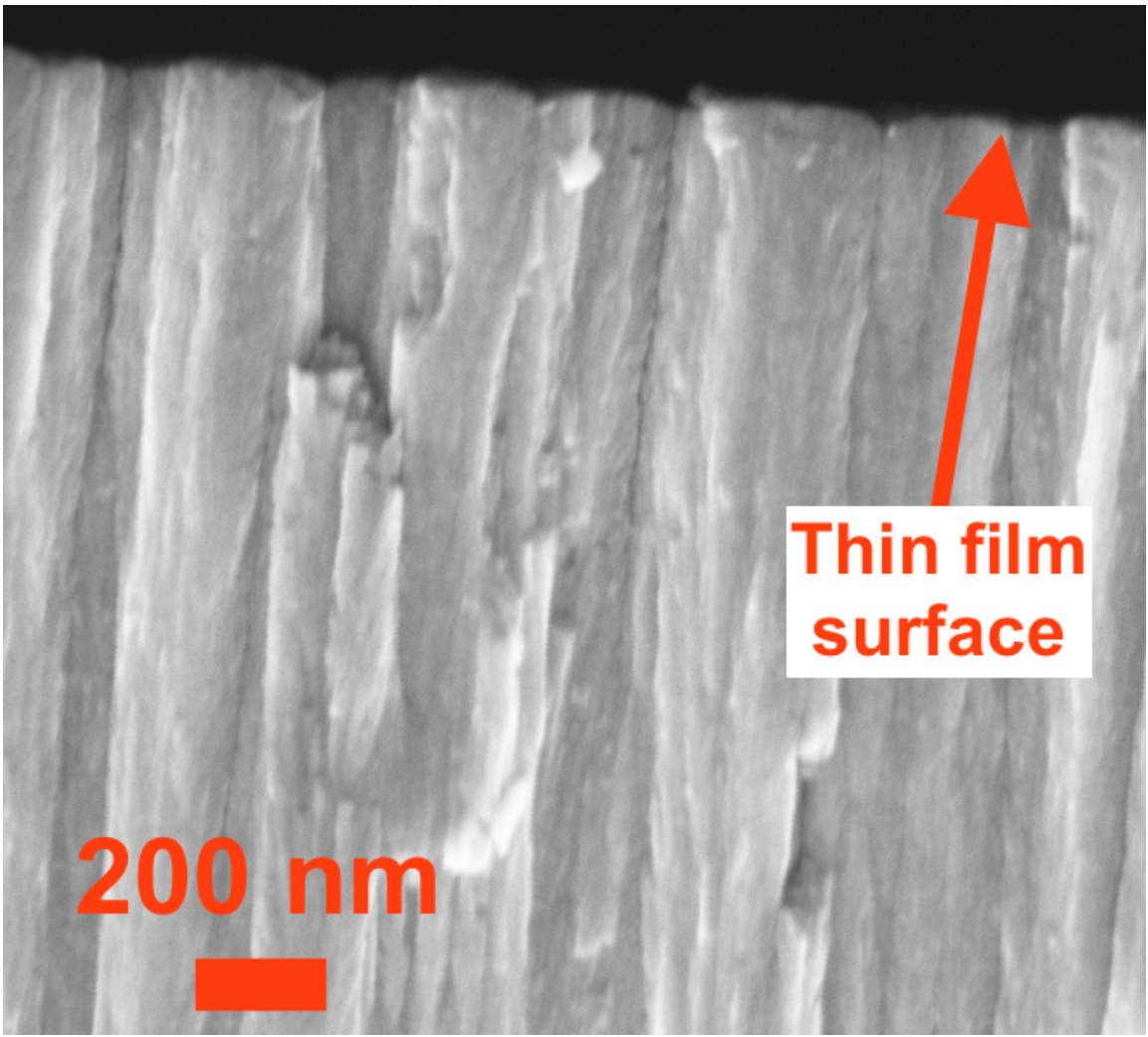

Figure 8. Cross-section SEM image of a-Si thin film

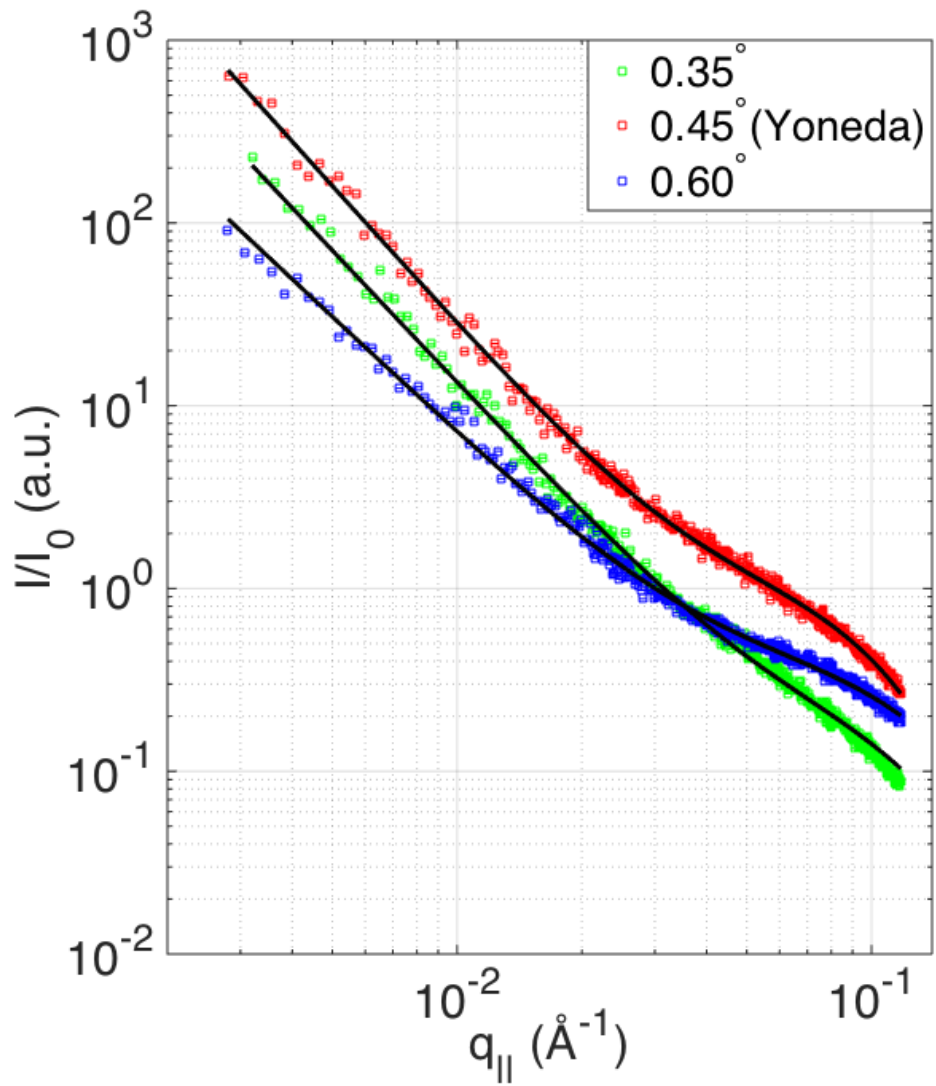

Figure 9. Steady-state GISAXS intensities measured at exit angles below, at, and above the Yoneda wing during a-WSi$_2$ thin film growth. The solid lines are fits to Eq. 9; fit parameters are given in Table 2.

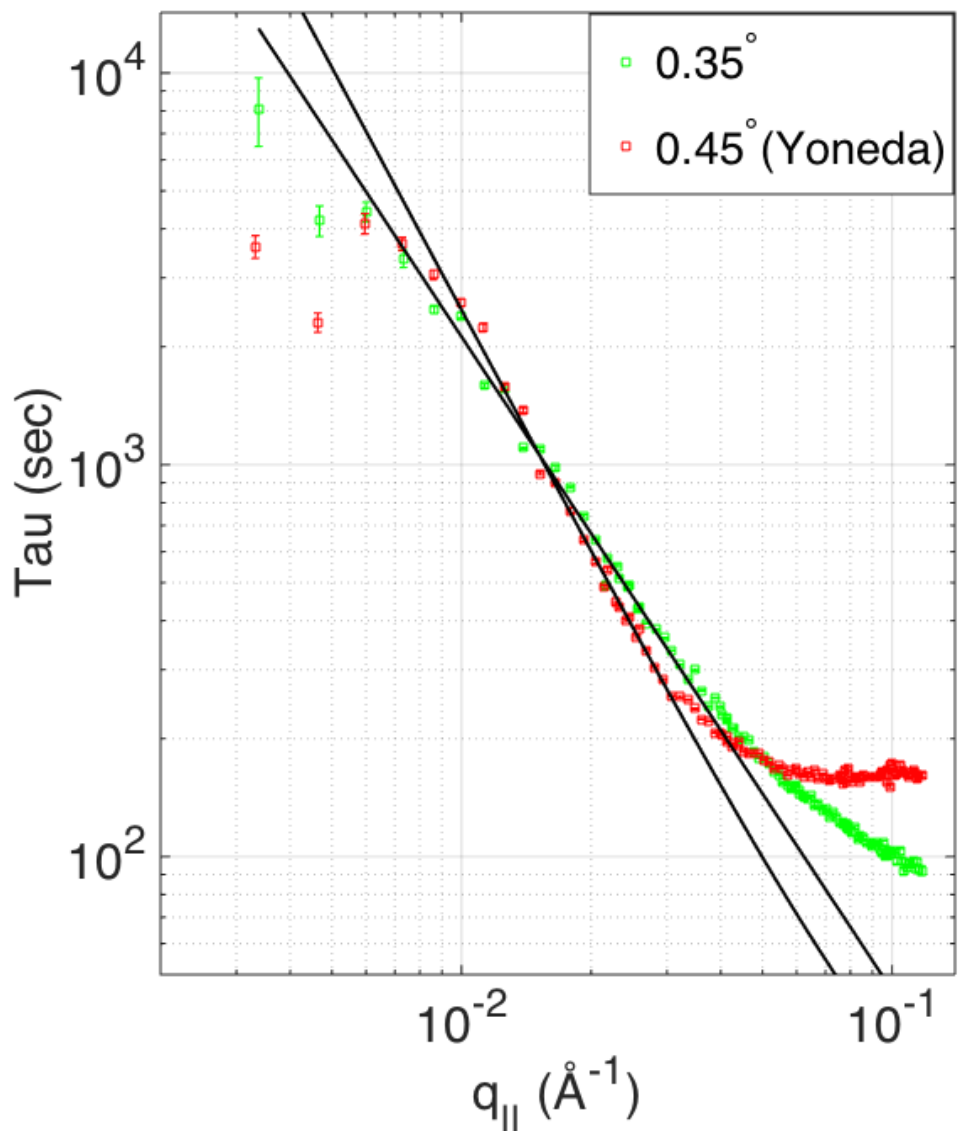

Figure 10. Correlation times for a-WSi$_2$ thin film growth measured at exit angles below and at the Yoneda wing.

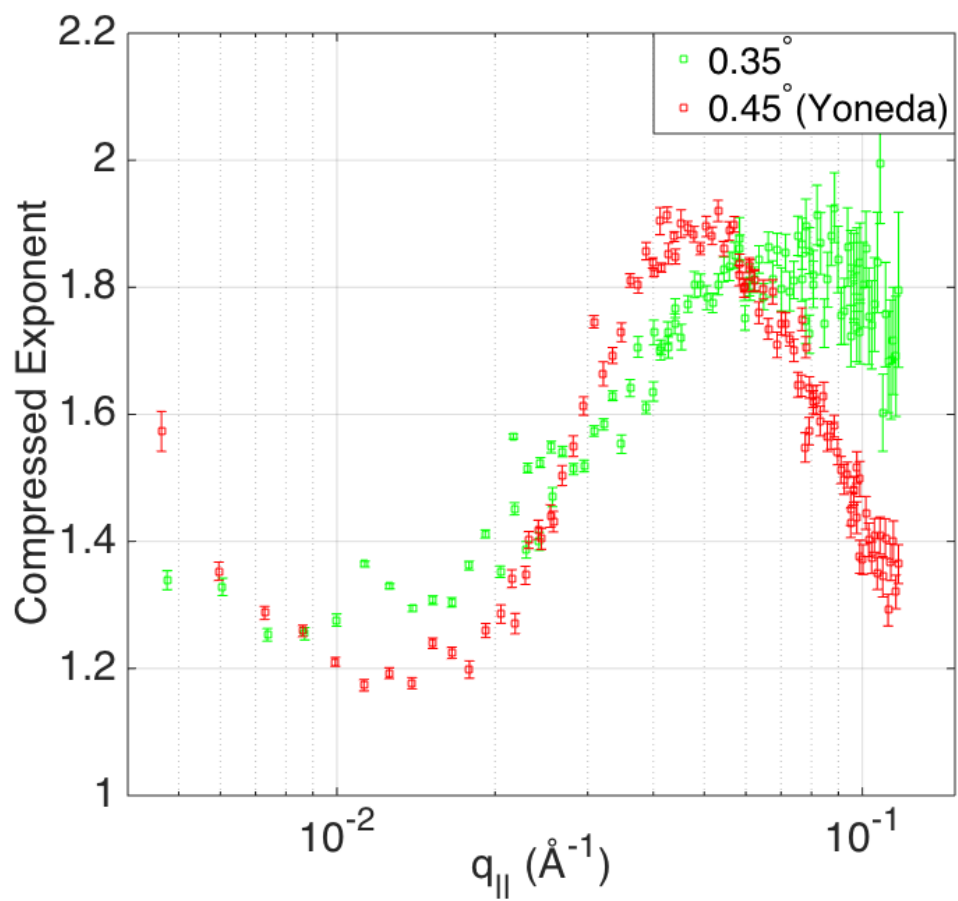

Figure 11. Compressed exponents from the $g_2(t)$ fits for exit angles below and at the Yoneda wing for a-WSi$_2$ growth.

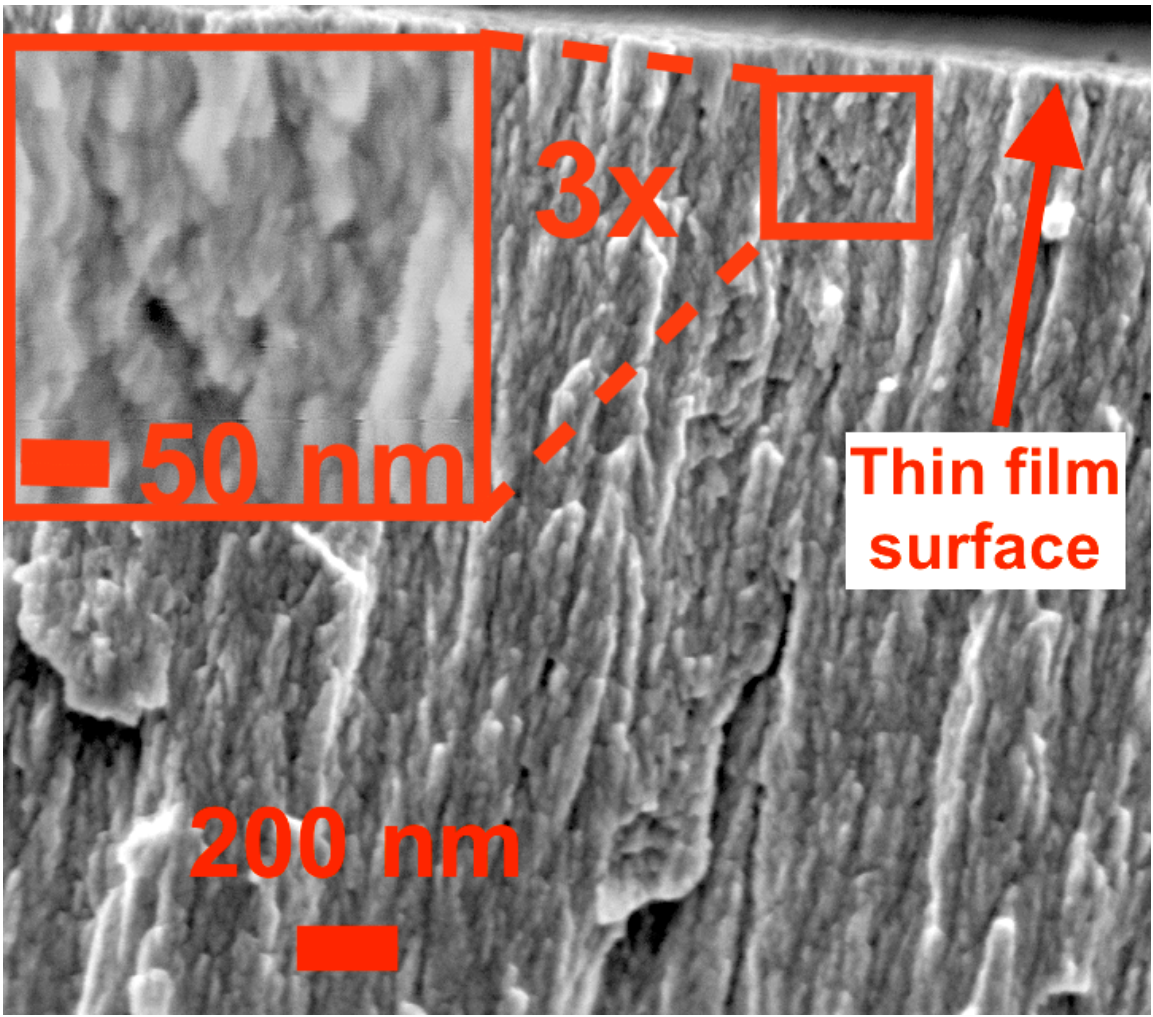

Figure 12. Cross-section SEM image of a-WSi$_2$ thin film